\documentclass[preprint,aps]{revtex4-1}
\usepackage{amssymb}
\usepackage{amsmath,amssymb,graphicx}
\usepackage{graphicx}
\usepackage{dcolumn}
\usepackage{bm}
\def\be{\begin{equation}}
\def\ee{\end{equation}}
\def\bea{\begin{eqnarray}}
\def\eea{\end{eqnarray}}

\def\lsim{\raise0.3ex\hbox{$\;<$\kern-0.75em\raise-1.1ex\hbox{$\sim\;$}}}
\def\gsim{\raise0.3ex\hbox{$\;>$\kern-0.75em\raise-1.1ex\hbox{$\sim\;$}}}

\def\eg{{\it e.g.}}
\def\ie{{\it i.e.}}
\newcommand{\paren}[1]{\left(#1\right)}
\newcommand{\pacua}[1]{\left[#1\right]}
\newcommand{\llav}[1]{\left\{#1\right\}}

\newcommand{\refeq}[1]{(\ref{#1})}

\newcommand{\Tr}{\mathop{\mathrm{Tr}}}

\newcommand{\Rea}{{\rm Re}\,}

\begin{document}

\title{Dynamical Symmetry Breaking With a Fourth Generation}%

\author{D. Delepine}
\email{delepine@fisica.ugto.mx}
\affiliation{Departamento de F\'isica,  Universidad de
Guanajuato, Campus Le\'on,
 C.P. 37150, Le\'on, Guanajuato, M\'exico.}
\author{M. Napsuciale}
\email{mauro@fisica.ugto.mx}
\affiliation{Departamento de F\'isica,  Universidad de
Guanajuato, Campus Le\'on,
 C.P. 37150, Le\'on, Guanajuato, M\'exico.}%
\author{C. A. Vaquera-Araujo}
\email{vaquera@fisica.ugto.mx}
\affiliation{Departamento de F\'isica,  Universidad de
Guanajuato, Campus Le\'on,
 C.P. 37150, Le\'on, Guanajuato, M\'exico.}

\date{\today}

\begin{abstract}
Adding a fourth generation to the Standard Model and assuming it to be valid 
up to some cutoff $\Lambda$, we show that electroweak symmetry is broken by radiative corrections due 
to the fourth generation.  The effects of the fourth generation are isolated using a 
Lagrangian with a genuine scalar without self-interactions at the classical level. For 
masses of the fourth generation consistent with electroweak precision data (including the 
$B\rightarrow K \pi$ CP asymmetries) we obtain a Higgs mass of the order of a few hundreds GeV 
and a cutoff $\Lambda$ around 1-2 TeV.  We study the reliability of the perturbative treatment used 
to obtain these results taking into account the running of the Yukawa couplings of the fourth quark 
generation with the aid of the Renormalization Group (RG) equations, finding similar allowed values for 
the Higgs mass but a slightly lower cut-off due to the breaking of the perturbative regime.  Such low 
cut-off means that the effects of new physics needed to describe electroweak interactions at  energy 
above $\Lambda$ should be measurable at the LHC.We use the minimal supersymmetric extension of the 
standard model with four generations as an explicit example of models realizing the dynamical electroweak 
symmetry breaking by radiative corrections and containing new physics. Here, the cutoff is replaced 
by the masses of the squarks and electroweak symmetry breaking by radiative corrections requires the 
squark masses to be of the order of 1 TeV. 
\end{abstract}

\maketitle

\section{Introduction.}

Many experimental results on $B$ physics (see \cite{acpref}) can be seen as hints of physics beyond 
the Standard Model (SM). Electroweak precision data also points to new physics scenarios 
\cite{Chanowitz:2001}. In the LHC era, new physics related to the observability of the Higgs 
boson is worthy to study and the elucidation of the Higgs sector properties is a topic of utmost importance.

A simple extension of the Standard Model (SM) is the introduction of a new generation of quarks and leptons (SM4). 
Precision data do not exclude the existence of a sequential fourth generation 
\cite{Maltoni:1999ta, He:2001tp, Novikov:2002tk, Kribs:2007nz, Hung:2007ak, Hashimoto:2010at}.
An extensive review and an exhaustive list of references to the work on the subject previous to our 
century can be found in \cite{Frampton:1999xi}. Recent highlights on consequences of a fourth 
generation can be found in \cite{Holdom:2009rf}. These include mechanisms of dynamical electroweak symmetry 
breaking by condensates of fourth generation quarks and leptons 
\cite{Holdom:1986rn,Hill:1990ge, Carpenter:1989ij, Hung:2009hy}, convergence improvement of the 
three SM gauge couplings due to the Yukawa coupling contributions from the fourth generation \cite{Hung:1997zj}, 
the possibility of electroweak baryogenesis through first-order electroweak phase transition with four generations 
\cite{Ham:2004xh, Fok:2008yg, Kikukawa:2009mu}, CP violation based on Jarlskog invariants generalized to 
four generations \cite{Hou:2008xd} and the hierarchy problem \cite{Hung:2009ia}.

The $B\to K \pi$ CP asymmetries puzzles can also be easily solved by a fourth generation 
\cite{Soni:2008bc,Hou:2005hd,Hou:2006jy} for a range of extra quark masses within the values allowed by  
high precision LEP measurements \cite{Maltoni:1999ta,He:2001tp,Novikov:2002tk,Bobrowski:2009ng}, namely 
\begin{eqnarray}\label{eq:benchmarks}
m_{\ell_4} - m_{\nu_4} &&\simeq  30 - 60 \; \mathrm{GeV}  \nonumber\\
m_{u_4} - m_{d_4}  &&\simeq
  \left( 1 + \frac{1}{5} \ln \frac{m_H}{115 \; \mathrm{GeV}}
                         \right) \times 50 \; \mathrm{GeV}  \\ 
|V_{u d_4}|,|V_{u_4 d}| &&\lsim 0.04 \ \nonumber\\
|U_{\ell_4}|,|U_{\mu_4}|   &&\lsim 0.02 \; , \nonumber
\end{eqnarray}
where $V$ ($U$) is the CKM (MNS) quark (lepton) mixing matrix which is now a $4\times 4$ unitary matrix. 
These bounds are subject to direct search limits from LEPII and CDF \cite{Achard:2001qw,Lister:2008is,Aaltonen:2009nr} :
\begin{eqnarray}\label{LEP-CDF}
m_{\nu_4,\ell_4} &> &  100\; \mathrm{GeV} \nonumber\\
m_{u_4} & >& 311\; \mathrm{GeV}\;   \\ 
m_{d_4} & >& 338\; \mathrm{GeV}.  \nonumber
\end{eqnarray}
In ref.\cite{Soni:2008bc,Hou:2006jy,Hou:2005hd}, in order to solve the CP asymmetry puzzles in 
$B\to K \pi$, one needs the extra quarks to be within the following range \cite{Soni:2008bc}:
\begin{eqnarray}
400\; \mathrm{ GeV} <& m_{u_4} & < 600\; \mathrm{ GeV} . 
\end{eqnarray}

Such values of new quark masses imply strong Yukawa couplings. So, it is natural to expect that 
this fourth generation could play a special role in the electroweak symmetry breaking (EWSB). Contrary to 
other works where it is assumed that Yukawa couplings are strong enough to produce composite scalars at 
low energy\cite{Holdom:1986rn,Hill:1990ge, Carpenter:1989ij, Hung:2009hy}, we shall assume that the 
perturbative treatment \cite{Coleman:1973jx} is still valid. This assumption is 
justified by the fact that even fourth generation masses in the range of 300-600 GeV imply 
Yukawa couplings ($g_{f}$) around 2-3. In the loop expansion, the perturbative parameters are given 
by $g_{f}^2/4 \pi$ which are still smaller than one for these mass values. 

In this work we study the effect of a fourth generation in the dynamical breaking of electroweak symmetry. In order 
to isolate these effects and following the spirit of \cite{Fatelo:1994qf}, we start in Section II with a model 
with vanishing scalar self-interactions at the classical level and maintain this condition at the one loop level. 
In this model the symmetry breaking of the gauge group $SU(2)_L\times U(1)_Y$  is a dynamical effect exhausted by the Yukawa couplings of chiral fermions to the Higgs scalar. In Section III we relax the condition of vanishing effective self-interactions and perform a Renormalization Group (RG) improvement in order to determine whether perturbative conditions remain valid when the running of Yukawa couplings are taken into account.  Finally, in Section IV we also explore the implications of this kind of dynamical EWSB mechanism in the minimal supersymmetric standard 
model (MSSM) extended with a fourth generation of chiral matter (MSSM4) (see \cite{Chaichian:1995ef} for a closely related approach) which has been studied in many situations \cite{Dubicki:2003am,Carena:1995ep,Gunion:1995tp}.

\section{ Symmetry breaking induced by the fourth generation}
We start with the Lagrangian describing electroweak interactions and consider only the part required for our 
purposes, namely  
\begin{equation}\label{lag1}
{\mathcal L}= \frac{1}{2}\partial^\mu\phi\partial_\mu\phi-\frac{\mu^2(v)}{2}\phi^2 -\frac{\lambda(v)}{4!}\phi^4
+\sum_{a}\pacua{\bar\psi^a i\gamma^\mu\partial_\mu\psi^a -\frac{g_{a}(v)}{\sqrt{2}}\phi \bar\psi^a \psi^a}. 
\end{equation}
Here, $\phi$ is the neutral component of the standard Higgs doublet and $\psi^{a}$ is the corresponding 
fermion field with $a=t,u_4,d_4,\ell_4,\nu_4$. We assume that our description of the electroweak interaction 
by the symmetries of the Standard Model is valid only up to a cutoff $\Lambda$, but our perturbative expansion 
will be done on the physical couplings at the scale $v$ (see e.g.\cite{Aitchison} for a discussion on this viewpoint), 
a fact that we emphasize by explicitly showing the dependence of the parameters on this scale which --anticipating 
results-- we identify below as the electroweak symmetry breaking scale.

As it is well known, if $ \mu^{2}(v)<0$ and 
$\lambda(v)>0$ we have spontaneous symmetry breaking (SSB) already at tree level.  In this model we are 
interested in the possibility of triggering EWSB without invoking a spontaneous breakdown, and therefore, we 
require an authentic scalar field, \ie,  $\mu^{2}(v)>0$. We expect SB to be induced by quantum effects 
and we are specially interested in the isolation of the effects due to the fourth generation in such a dynamical EWSB. 
With this aim, we start taking $\lambda(v)=0$, which is the limiting case where one-loop effects of the 
scalar sector are completely suppressed and only the matter sector is responsible for EWSB.  The condition
$\lambda(v)=0$ should not be taken as a fundamental requirement of the model nor as a fine tunning condition,
but instead as the limiting scenario where the effects of the fourth generation are more easily recognizable. 

The one-loop corrections to the classical potential $V^{(0)}=\frac{1}{2}\mu^2(v)\phi^2$ can be calculated 
using standard techniques \cite{Sher:1988mj}. At one loop level we obtain  
\begin{equation}
V^{(1)}_{f}=\frac{\mu^2(v)}{2}\phi^2-\sum_{a} \frac{4N_c^a}{32\pi^2}\int_0^{\Lambda^2} dk_E^2 k_E^2  
\ln\left[\frac{k_E^2+ m_a^2(\phi)}{k_E^2+ m_a^2(0)}\right],
\end{equation}
where $N_c^a$ is the number of colors of the field labeled by $a$ and $m_a^2(\phi)=g_{a}^2(v)\phi^2/2$. 
Including one-loop gauge boson contributions to this potential is straightforward and yields
\begin{equation}\label{Vef}
\begin{split}
V^{(1)}=&\frac{\mu^2(v)\phi^2}{2}+\sum_{a} \frac{n_a}{32\pi^2}\int_0^{\Lambda^2} dk_E^2 k_E^2 
\ln\left[\frac{k_E^2+ m_a^2(\phi)}{k_E^2+ m_a^2(0)}\right]\\
=&\frac{\mu^2(v)\phi^2}{2}+\sum_{a} \frac{n_a}{64\pi^2}\biggl\{ \pacua{m_a^2(\phi)-m_a^2(0)}\Lambda^2
+\Lambda^4 \ln\left[ \frac{\Lambda^2+m_a^2(\phi)}{\Lambda^2+m_a^2(0)}\right]
\\&\qquad\qquad-m_a^4(\phi)\ln\left[1+ \frac{\Lambda^2}{m_a^2(\phi)}\right]+m_a^4(0)
\ln\left[1+ \frac{\Lambda^2}{m_a^2(0)}\right]\biggr\}
\end{split}
\end{equation}
where now $a=t,u_4,d_4,\ell_4,\nu_4, W, Z$ and the field-dependent squared masses for gauge bosons are 
given by $m_W^2(\phi)=g_2^2\phi^2/4$ and $m_Z^2(\phi)=(g_1^2+g_2^2)\phi^2/4$, with $g_{1}$ and $g_{2}$ 
as the $U(1)$ and $SU(2)$ gauge couplings evaluated at the scale $v$ respectively. Consequently, 
the degeneracies per particle are the following: $n_W=6$, $n_Z=3$, $n_t=n_{u_4}=n_{d_4}=-12$ and $n_{\ell_4}=n_{\nu_4}=-4$.

\begin{center}
\begin{figure}[ptb]
\begin{center}
\includegraphics[width=0.8 \textwidth]{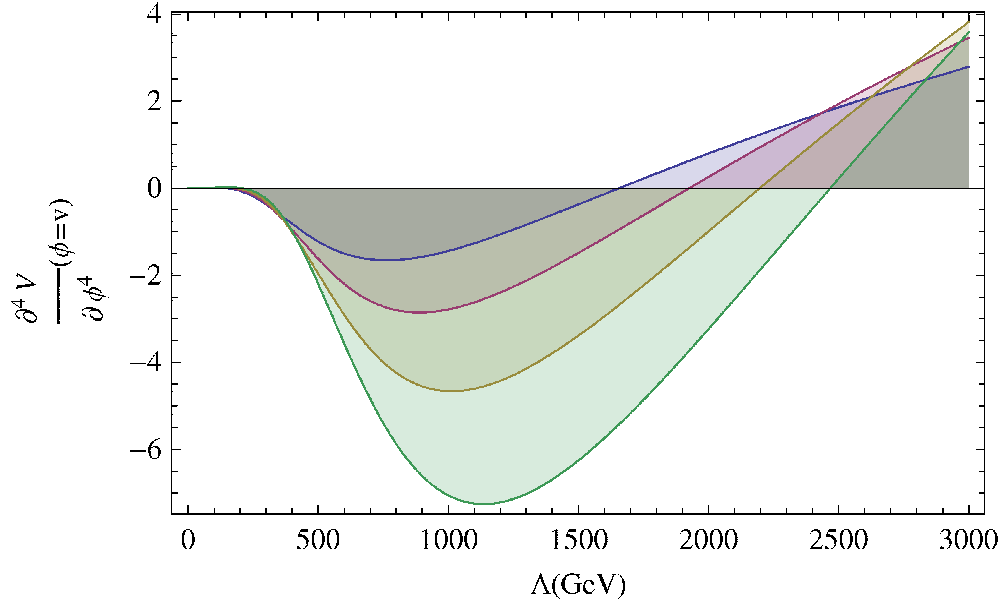}
\end{center}
\caption{Effective Higgs self-coupling at the electroweak scale $v=246~$GeV as a function of the cutoff $\Lambda$ 
for fixed values of the heavy fermions. The curves correspond to $m_{u_4}=$350, 400, 450 and 500 GeV 
with $m_{\ell_4}=200$ GeV and mass splittings $m_{u_4} - m_{d_4}  =60$ GeV and $ m_{\ell_4} - m_{\nu_4} =45 $ GeV from shallowest to deepest.}
\label{lambda}
\end{figure}
\end{center}

From \refeq{Vef}, one can see that the classical minimum $\langle\phi\rangle=0$ can be turned into a local 
maximum by the one-loop corrections. A new minimum appears then at $\langle\phi\rangle=v\neq 0$ and all 
particles in the model acquire a mass $m_a=m_a(v)$. The only non-trivial solution to 
$\partial V^{(1)}/\partial\phi |_{\phi=v}=0$  is
\begin{equation}
 \mu^2(v)=-\sum_{a}\frac{n_a m_a^4}{16\pi^2 v^2}\pacua{\frac{\Lambda^2}{m_a^2}-\ln\paren{1+\frac{\Lambda^2}{m_a^2}}},
\end{equation}
with $\mu^2(v)>0$ for the inputs of the problem as required, meaning that the tree level scalar mass 
term is genuine and symmetry breaking is entirely driven by one-loop effects. The Higgs boson mass at one 
loop level can be identified as
\begin{equation}\label{HM}
m_H^2(v)=\left.\frac{\partial^2 V^{(1)}}{\partial\phi^2}\right|_{\phi=v}=
-\sum_{a}\frac{n_a m_a^4}{8\pi^2v^2}\pacua{\ln\paren{1+\frac{\Lambda^2}{m_a^2}}-\frac{\Lambda^2}{m_a^2+\Lambda^2}}
\end{equation}
and the fourth derivative of the effective potential evaluated at the scale $v$ reads
\begin{equation}\label{cuarta}
\left.\frac{\partial^4 V^{(1)}}{\partial\phi^4}\right|_{\phi=v}=
-\sum_{a}\frac{3 n_a m_a^4}{8\pi^2v^4}\pacua{\ln\paren{1+\frac{\Lambda^2}{m_a^2}}+9\frac{m_a^2}{m_a^2+\Lambda^2}
-8\frac{m_a^4}{(m_a^2+\Lambda^2)^2}+\frac{8}{3}\frac{m_a^6}{(m_a^2+\Lambda^2)^3}-\frac{11}{3}}.
\end{equation}
The effective scalar self-interaction depends on the fermion masses $m_{a}$, the minimum of the 
effective potential $v$ and the cut-off $\Lambda$ and it is worthy to study this dependence. This is shown in 
Fig. (\ref{lambda}) for $v=246 ~$GeV  and heavy fermion masses in the range given in  
Eqs.(\ref{eq:benchmarks}, \ref{LEP-CDF}). 

Notice that for given fermion masses, the specific value of the effective self-interaction at the 
electroweak symmetry breaking scale depends on the value of the unknown scale $\Lambda$. Up to this point, 
a wide range of possible values for the cut-off are eligible and one must take into account the dependence 
of the parameters of the model on $\Lambda$, as we will do in section III. However, these values must be 
consistent with the perturbative treatment we are using which requires a small effective scalar 
self-interaction. From Fig. (\ref{lambda}) we can see that this narrows the range of values for $\Lambda$, 
the allowed range depending on the specific masses of the fourth generation. Interestingly, 
for given values of the fermion masses, there are specific values of $\Lambda$ such that the effective 
self interaction also vanishes. These specific values are worthy to study 
in detail because in this case the  effects of scalar self-interactions in the EWSB at the next order in 
perturbation theory also vanish and EWSB is still driven by the Yukawa couplings at that order. Furthermore, 
in this case the scale $\Lambda$ is fixed by the electroweak scale $v$ and the values of the fermion masses.

There are two solutions to the equation 
\begin{equation}\label{cuarta0}
\left.\frac{\partial^4 V^{(1)}}{\partial\phi^4}\right|_{\phi=v}=0,
\end{equation}
for heavy fermion masses in the range given in Eqs.(\ref{eq:benchmarks}, \ref{LEP-CDF}). One of them yields 
$\Lambda$ around the electroweak symmetry breaking scale $v$ and we consider it as unphysical. The other 
solution lies in the range 
\begin{equation}\label{range}
1600\text{ GeV}<\Lambda<2500\text{ GeV},
\end{equation}
depending on the input for the masses of the fourth generation fermions.

\begin{center}
\begin{figure}[ptb]
\begin{center}
\includegraphics[width=0.8 \textwidth]{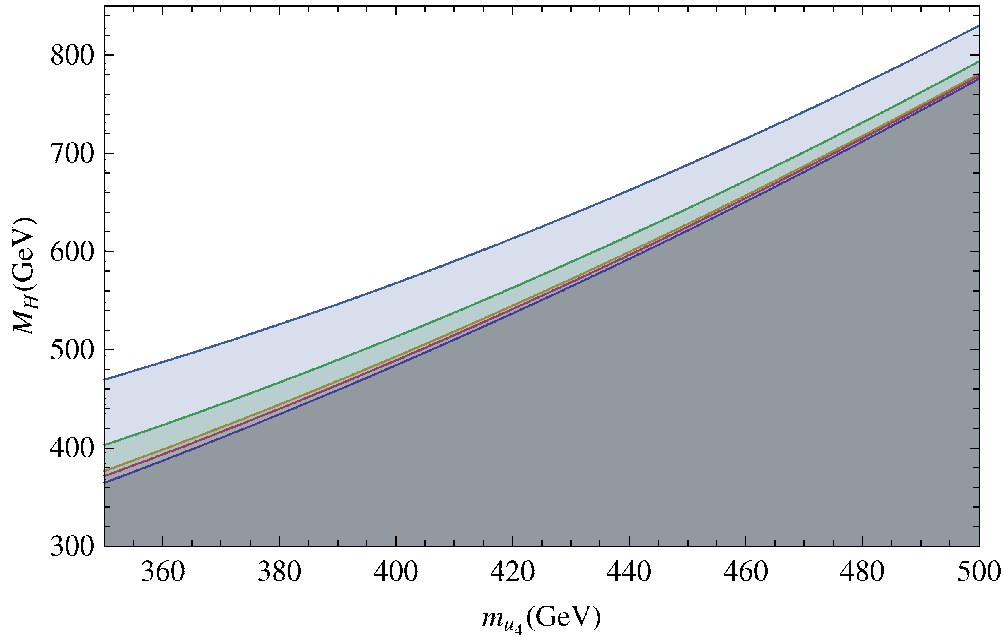}
\end{center}
\caption{Higgs mass as a function of $m_{u_4}$ for different values of  $m_{\ell_4}=400,~300,~200,~100 $ GeV from top to bottom. The lowest line contains only the contribution of $u_4$ and $d_4$.}
\label{mu4}
\end{figure}
\end{center}

\begin{center}
\begin{figure}[ptb]
\begin{center}
\includegraphics[width=0.8 \textwidth]{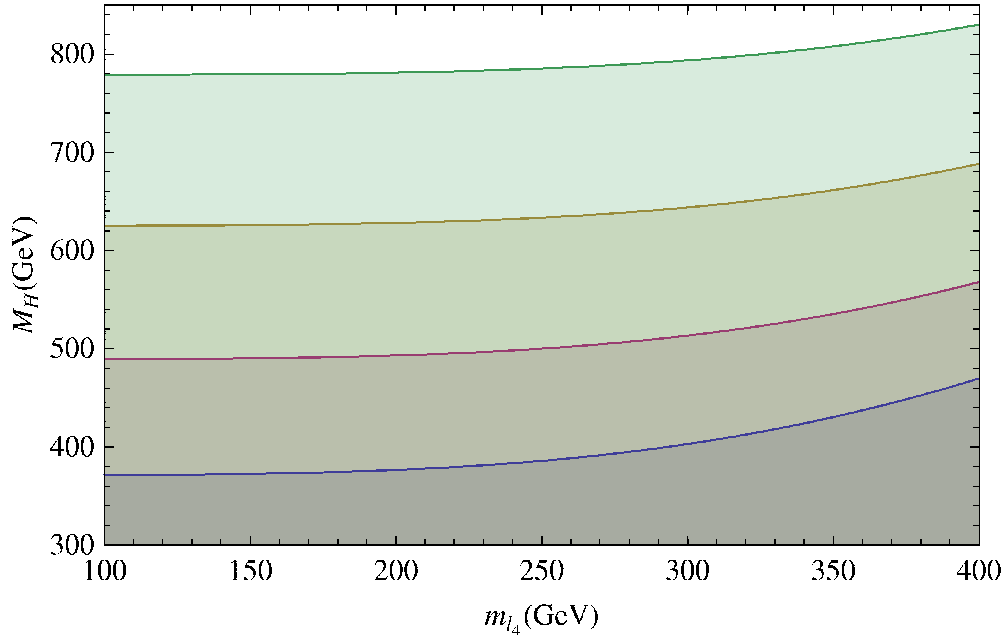}
\end{center}
\caption{Higgs mass as a function of $m_{\ell_4}$ for different values of $m_{u_4}=350,~400,~450,~500$ GeV from bottom to top.}
\label{ml4}
\end{figure}
\end{center}

Once we have fixed the cutoff $\Lambda$ for given masses of the heavy fermions, we obtain from Eq. 
\refeq{HM} the corresponding Higgs mass as a function of the fourth generation quark and lepton masses. 
In the numerical analysis we use
\begin {equation}\label{dif}
\begin{array}{ccc}
 m_{\ell_4} - m_{\nu_4} =45 \; \mathrm{GeV},&\qquad & 100 \; \mathrm{GeV} \le m_{\ell_4}\le 400 \; \mathrm{GeV},  \\
m_{u_4} - m_{d_4}  =60 \; \mathrm{GeV},&\qquad & 350 \; \mathrm{GeV} \le m_{u_4}\le 500 \; \mathrm{GeV},
\end{array}
\end {equation}
as suggested by Eqs.(\ref{eq:benchmarks}, \ref{LEP-CDF}). Under these considerations, the Higgs mass is a 
smooth function of $m_{u_4}$ and $m_{\ell_4}$ and has a more pronounced dependence on $m_{u_4}$ as shown 
in Figs.(\ref{mu4}, \ref{ml4}). More important, a modest Higgs mass of $\sim 350 $ GeV is 
reachable and even a heavy Higgs of $\sim 800 $ GeV would be consistent with electroweak 
precision data if EWSB is entirely driven by Yukawa forces of the hypothetical fourth generation and the top quark.
Notice that if fourth generation 
lepton masses are of order $100-200$ GeV, then the contribution of $u_4$ and $d_4$ almost determine 
completely the Higgs mass prediction, as depicted in Fig.(\ref{mu4}). Combining Eq.\refeq{HM} and 
Eq.\refeq{lambda} with condition Eq.\refeq{cuarta0}, this fact is expressed as follows:
\begin{equation}\label{HM1}
m_H^2\approx\sum_{q=u_4,d_4}\frac{4 m_q^4}{\pi^2v^2}\pacua{1-3\frac{m_q^2}{\Lambda^2}
+\mathcal{O}\paren{\frac{m_q^4}{\Lambda^4}}}.
\end{equation}
Also in this case, the cutoff for new physics should be within the range given in Eq.\refeq{range}.  
A simple and good approximation for this case is 
\begin{equation}\label{HMa}
m_H\approx\frac{1.8
9}{\pi v}\sqrt{m_{u_4}^4+m_{d_4}^4},
\end{equation}
with $\Lambda\approx 5 m_{u_4}$.

It is important to remark that even for masses of the 4th generation around 500 GeV, the corresponding Yukawa 
couplings ($g_{q_4}$) are around $2-3$ thus the loop expansion parameter, given by $g_{q_4}^2/4\pi$, 
is smaller than one and justifies our perturbative approach. 

Results contained in Figs.(\ref{mu4}, \ref{ml4}) are in agreement --{\it mutatis mutandis}-- with the analysis performed in the full renormalized SM4 framework \cite{Hashimoto:2010at}.

\section{RG Improved Model}
It is important to check the consistency and stability of our previous  approach to take into account the 
running of the Yukawa couplings as it is well known that these couplings could reach the non perturbative 
regime very quickly. 
In this section, following the approach of \cite{Cvetic:1995va} we investigate the 
leading effects of the heavy fourth generation quarks on the scalar sector with a special emphasis on the 
perturbative nature of the analysis and its implications on the possible choices for the ultraviolet cut-off. 
We use the Renormalization Group equation (RG) to estimate the running of the couplings. 

From the previous results, we learned that the contribution of gauge bosons, top quark and fourth generation 
leptons to the one-loop effective potential is negligible compared to that of the fourth generation quarks 
if we assume that new leptons are relatively light. In a first approximation we will consider only the 
effects of the running in the fourth family of quarks. In order to incorporate these effects properly in 
the analysis of the Higgs mass, we will use the pole mass for the Higgs. Furthermore, the study of the perturbative regime will require the running of the fermion masses which are dictated by the running of 
the Yukawa couplings.  Since we will study the behavior of our observables as a function of these masses 
it is important to work with the fermion masses as defined at the corresponding scale, i.e.
$m_{q_{4}}=m_{q_{4}}(\mu=m_{q_{4}})$. Finally we will incorporate renormalization effects of the vacuum expectation value of the scalar field. All these effects are more easily handled using a more conventional approach thus, unlike the previous section, here we start with the bare Lagrangian whose  
sector of our primary interest is 
\begin{equation}\label{Lagr}
{\mathcal L}^{(\Lambda)}= \frac{1}{2}\partial^\mu\phi\partial_\mu\phi-V^{(0)} \left( \phi^2; \Lambda \right)
+\sum_{a=u_4,d_4}\pacua{\bar\psi^a i\gamma^\mu\partial_\mu\psi^a -\frac{g_a\left( \Lambda \right)}{\sqrt{2}}\phi \bar\psi^a \psi^a},
\end{equation}
with
 \begin{equation}
V^{(0)}\left( \phi^2; \Lambda \right)
= \frac{1}{2} \mu^2(\Lambda) \phi^2
+ \frac{\lambda\left( \Lambda \right) }{4!}
  \phi^4 \ .
\label{Vtree}
\end{equation}
At one loop level we have 
\begin{equation}
V^{(1)}\left( \phi^2; \Lambda \right)=V^{(0)}\left( \phi^2; \Lambda \right)-\sum_{a=u_4,d_4}\frac{4N_c^a}{32\pi^2}\int_0^{\Lambda^2} dk_E^2 k_E^2  
\ln\left[1 + \frac{ g^2_a(\Lambda) \phi^2}{2 k_E^2}\right].
\end{equation}
Again, if we insist in a dynamical SB triggered by fourth generation quarks and we set $\lambda(\Lambda)=0$, the only non-trivial solution to 
$\partial V^{(1)}/\partial\phi |_{\phi=\langle \phi \rangle_{1}}=0$  is
\begin{equation}
 \mu^2(\Lambda)=  \sum_{a=u_4,d_4}\frac{ g_a^2(\Lambda) N_{c}}
{8 \pi^2} \left[ \Lambda^2 -  m_a^{(0)2}(\Lambda)  
\ln \left( \frac{\Lambda^2}{ m_a^{(0)2}(\Lambda)} + 1 \right) \right] ,
\end{equation}
with $\mu^2(\Lambda)>0$ for the inputs of the analysis. This means that we have an authentic scalar in the SB sector. Here 
\begin{equation}
m_a^{(0)}(\Lambda) = 
\frac{g_a(\Lambda)  \langle \phi \rangle_{1} }{\sqrt{2}}
\label{1ltgapdef}
\end{equation}
with $\langle \phi \rangle_{1}$ as the ``bare'' vacuum expectation value, where the subscript denotes the fact that this is
an approximation with only the one-loop fourth generation quantum effects.

It is important to notice that our Lagrangian depends now on the values of the coupling at the cut-off scale. In 
this section {\it the cut-off scale will be defined as the scale where the perturbative regime for the Yukawa 
couplings is still valid}. Above this scale, the Yukawa couplings could get strong enough to generate 
non-perturbative effects as condensate formation or others.

In order to obtain predictions on physical quantities, we must make an adequate choice of $\Lambda$ taking special care in the preservation of the perturbative expansion.
The relations between the bare parameters of the model and the physical parameters proceed as follows: The physical (pole) mass $M_H$ of the scalar can be expressed in terms of the effective potential as
\begin{eqnarray}\label{MassH}
 M_H^2 & = &
\frac{ d^2 V^{(0)} }{d \phi^2} {\Big|}_
{\phi= \langle \phi \rangle_{1}} + \Sigma_{HH}\left(
q^2 =  M_H^2 \right)
\nonumber\\
& = & 
\frac{d^2 V^{(1)} }
{d \phi^2} {\Big|}_
{\phi= \langle \phi \rangle_{1}} - \Sigma_{HH}\left( q^2 = 0 \right)+ \Sigma_{HH}\left(
q^2 = M_H^2  \right),
\label{MHpole1}
\end{eqnarray}
where $\Sigma_{HH}( q^2 )$ stands for the scalar self energy, that can be approximated as the following truncated Green function calculated with fourth generation one-loop effects only
\begin{equation}
\begin{split}
-i \Sigma_{HH}( q^2 )&=-i \Sigma^{u_4u_4}_{HH}( q^2 )-i \Sigma^{d_4d_4}_{HH}( q^2 )\\  
&= \sum_{a=u_4,d_4}\left( \frac{g_a(\Lambda)}{\sqrt{2}} \right)^2
N_{c} \int \frac{d^4 k}{(2\pi)^4}
\Tr \left[ 
\frac{i}{\left( {k \llap /} - m_a^{(0)}(\Lambda) \right)}
\frac{i}{\left( {k \llap /} + {q \llap /} - m_a^{(0)}(\Lambda) \right)}
\right].                                  
\end{split}
\end{equation}
A straightforward calculation performing the Wick rotation in Euclidean space and imposing a spherical cut-off on the euclidean quark momentum yields
\begin{equation}\label{SHH}
\begin{split}
\Sigma_{HH}(q^2)=& -\sum_{a=u_4,d_4} \frac{ g_a^2(\Lambda) N_c}{8 \pi^2}\left\{\Lambda^2+\left[\frac{q^2}{2}-3m^{(0)2}_a(\Lambda)\right]\ln\left(\frac{\Lambda^2}{ m_a^{(0)2}(\Lambda)}\right)\right.\\
&+2m^{(0)2}_a(\Lambda)-\frac{7}{12}q^2+\frac{ m_a^{(0)2}(\Lambda)}{\Lambda^2}\left[\frac{q^2}{2}-5m^{(0)2}_a(\Lambda)\right]\\
&\left.+\frac{ m_a^{(0)4}(\Lambda)}{\Lambda^4}\left[q^2+\frac{7}{2}m^{(0)2}_a(\Lambda)\right]+\mathcal{O}\left((q^2;m_a^{(0)2}(\Lambda))\frac{ m_a^{(0)6}(\Lambda)}{\Lambda^2}\right)\right\}. 
\end{split}
\end{equation}

In this framework, the relation among the bare VEV $ \langle \phi \rangle_{1}$ and its renormalized counterpart $v\equiv\phi_{\text{ren}}$ is given by the renormalization of the kinetic scalar term and can be written as
\begin{equation}
Z_{\phi} \langle \phi \rangle_{1}^2= v^2,
\end{equation}
where
\begin{equation}\label{phiren}
\begin{split}
  Z_{\phi}=& \left[ 1 - \frac{d \Sigma_{HH}(q^2)}{d q^2}
{\Big|}_{q^2=M_H^2} \right]\\
 =& 1+\sum_{a=u_4,d_4} \frac{4  g_a^2(\Lambda) N_c}{64 \pi^2}\left\{\ln\left(\frac{\Lambda^2}{ m_a^{(0)2}(\Lambda)}\right)-\frac{7}{6}\right.\\
&\left.+\frac{ m_a^{(0)2}(\Lambda)}{\Lambda^2}+2\frac{ m_a^{(0)4}(\Lambda)}{\Lambda^4}+\mathcal{O}\left(\frac{ m_a^{(0)6}(\Lambda)}{\Lambda^6}\right)\right\}. 
\end{split}
\end{equation}

In the matter sector, the running of the relevant Yukawa couplings can be summarized in the following Renormalization Group Equations:
\begin{eqnarray}
(16\pi^2)\mu\frac{\partial}{\partial\mu}g_{u_4} =\frac{9}{2}g^3_{u_4}+\frac{3}{2}g_{u_4}g^2_{d_4}\\
(16\pi^2)\mu\frac{\partial}{\partial\mu}g_{d_4} =\frac{9}{2}g^3_{d_4}+\frac{3}{2}g_{d_4}g^2_{u_4}.
\end{eqnarray}
In the approximation $g_{u_4}\approx g_{d_4}$, defining $g_{u_4}-g_{d_4}\equiv\Delta g $, the previous equations reduce to
\begin{eqnarray}
(16\pi^2)\mu\frac{\partial}{\partial\mu}g_{u_4} \approx 6g^3_{u_4}\\
(16\pi^2)\mu\frac{\partial}{\partial\mu}\Delta g \approx 12 g^2_{u_4}\Delta g
\end{eqnarray}
and the solution can be written as
\begin{equation} g_{u_4}(\mu)\approx\left[\frac{1}{g^2_{u_4}(\mu_0)}-\frac{6}{16\pi^2}\ln\left(\frac{\mu^2}{\mu_0^2}\right)\right]^{-1/2}
\end{equation}
\begin{equation}
 \Delta g(\mu)\approx\Delta g(\mu_0)\left[1-\frac{3 g^2_{u_4}(\mu_0)}{8\pi^2}\ln\left(\frac{\mu^2}{\mu_0^2}\right)\right]^{-1}.
\end{equation}
The physical mass of the heaviest fourth generation quark are defined as 
\begin{equation}
 m_{u_4}\equiv
\frac{g_{u_4}(m_{u_4}) v}{\sqrt{2}}.
\end{equation}
The running of Yukawa couplings from $E=m_{u_4}$ to $E'=\Lambda$ is given by
\begin{equation}\label{g4} g_{u_4}(\Lambda)\approx\left[\frac{1}{g^2_{u_4}(m_{u_4})}-\frac{6}{16\pi^2}\ln\left(\frac{\Lambda^2}{m_{u_4}^2}\right)\right]^{-1/2}
\end{equation}
\begin{equation}
 \Delta g(\Lambda)\approx\Delta g(m_{u_4})\left[1-\frac{3 g^2_{u_4}(m_{u_4})}{8\pi^2}\ln\left(\frac{\Lambda^2}{m_{u_4}^2}\right)\right]^{-1}.
\end{equation}

Inserting \refeq{SHH} into \refeq{MassH}, the squared pole mass of the scalar is
\begin{eqnarray}
 M_H^{2} & = & \sum_{a=u_4,d_4} \frac{8  g_a^2(\Lambda) N_c Z_{\phi}^{-2}v^2}{64 \pi^2}\left[\ln\left(\frac{\Lambda^2}{ m_a^{(0)2}(\Lambda)}+1\right)-\frac{ \Lambda^2}{\Lambda^2+m_a^{(0)2}(\Lambda)}\right]. 
\label{MHpole2}
\end{eqnarray}
In this case, maximum cut-off can be naturally defined as the largest scale at which the model remains perturbative. That scale is achieved when alpha-Yukawa becomes equal to one
\begin{equation}\label{pert}
 \frac{g^2_{u_4}(\Lambda_\text{max})}{4\pi}=1,
\end{equation}
which can be solved to yield
\begin{equation}\label{lmax}
\Lambda_{\text{max}}=m_{u_4} e^{\frac{2\pi^{2}v^{2}}{3m_{u_4}^2}(1-\frac{m_{u_4}^{2}}{2\pi v^2})}
\end{equation}

\begin{center}
\begin{figure}[ptb]
\begin{center}
\includegraphics[width=0.8 \textwidth]{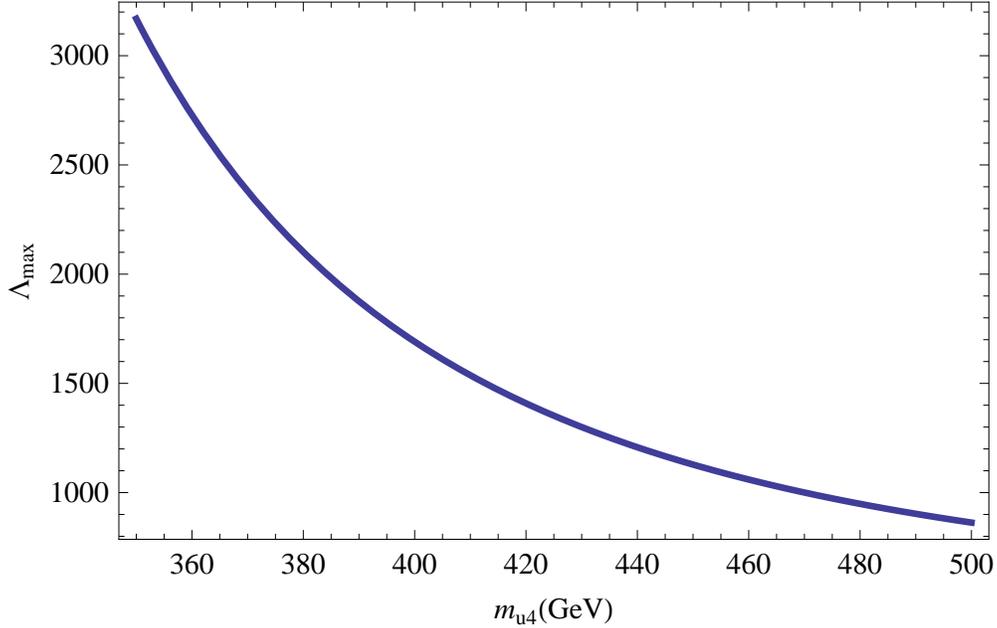}
\end{center}
\caption{$\Lambda_\text{max}$ as a function of the physical mass of the heaviest quark  $m_{u_4}$}
\label{L00}
\end{figure}
\end{center}

In Fig.(\ref{L00}), $\Lambda_\text{max}$ is shown as a function of the heaviest quark mass. Hence, for a quark with mass $m_{u_4}=400$ GeV the maximum cut-off is $\Lambda_\text{max}\approx 1700$ GeV; while for $m_{u_4}=500$ GeV we have $\Lambda_\text{max}\approx 860$ GeV. Even in this case, a Higgs mass between $350$ GeV and $650$ GeV is compatible with fourth generation quarks with masses between $350$ GeV and $500$ GeV which are responsible for the EWSB in a perturbative fashion with a physical cut-off $\Lambda<\Lambda_\text{max}$. 

\begin{center}
\begin{figure}[ptb]
\includegraphics[width= 0.8 \textwidth]{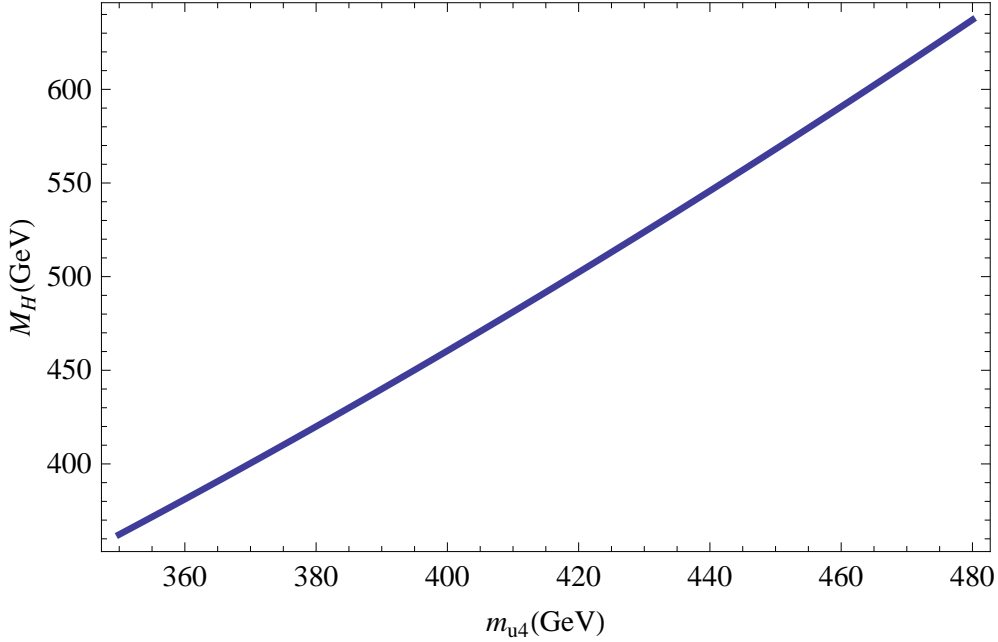}
\caption{Higgs (pole) mass as a function of $m_{u_4}$ with $\Lambda=2m_{u_4}$ for $\Delta m=m_{u_4}-m_{d_4}=60$ GeV. }
\label{0060}
\end{figure}
\end{center}

Comparing with the previous section we can see that perturbative effects indeed appear at a lower scale than 
the naive scale for new physics. Still, taking the worst case, \eg, $\Lambda=2 m_{u_4}$, the predicted Higgs mass lies in the same range as before and the perturbative expansion is valid up to $m_{u_4}\approx 480$GeV, where $\Lambda\approx\Lambda_\text{max}$.  This is shown explicitly in Fig.(\ref{0060}), where the curve represents the  Higgs mass that correspond to the cut-off choice $\Lambda=2m_{u_4}$ as a function of the mass of the heaviest quark $m_{u_4}$ with $\Delta m=m_{u_4}-m_{d_4}=60$ GeV. Thus, the predictions 
of the previous model are not strongly modified by the RG Yukawa couplings, but the interpretation of the cutoff scale is different.

\section{Dynamical symmetry breaking in MSSM4}
We now perform an analogous calculation in the context of a low energy supersymmetric extension of the 
SM with a fourth generation of chiral matter. As is well known, in the Higgs sector of MSSM there are 
two scalar doublets of opposite hypercharge: $H_d=(H_d^0,H_d^-)^T$, $H_u=(H_u^+,H_u^0)^T$. Breaking 
supersymmetry softly, the tree-level scalar potential for the CP-even neutral scalars $H_1\equiv\Rea H_d$ 
and $H_2\equiv\Rea H_u$ is 
\begin{equation}\label{V00}
 V^{(0)}=\frac{1}{2}\left(H_1 H_2\right)\left(\begin{array}{cc}
                           m_1^2 & -m_{12}^2\\
			  -m_{12}^2 & m_2^2
                          \end{array}
\right)
\left(\begin{array}{c}
                           H_1\\
			  H_2
                          \end{array}\right)
+\frac{(g_1^2+g_2^2)}{32}(H_2^2-H_1^2)^2.
\end{equation}
The linear combination
\begin{equation}\label{red}
\left(\begin{array}{c}
                          \phi\\
			  \varphi
                          \end{array}\right)=\left(\begin{array}{cc}
                           \cos\beta & \sin\beta\\
			  -\sin\beta & \cos\beta
                          \end{array}
\right)
\left(\begin{array}{c}
                           H_1\\
			  H_2
                          \end{array}\right)
\end{equation}
with $\tan2\beta=2m_{12}^2/(m_2^2-m_1^2)$ diagonalizes the mass matrix in Eq.\refeq{V00} and the potential becomes
\begin{equation}\label{V002}
 V^{(0)}=\frac{\mu^2}{2}\phi^2+\frac{M^2}{2}\varphi^2
+\frac{(g_1^2+g_2^2)}{32}\left[ \cos2\beta(\varphi^2-\phi^2)+\sin2\beta \phi\varphi \right]^2,
\end{equation}
where
\begin{equation}\label{mum}
 \mu^2,M^2=\frac{1}{2}\pacua{m_1^2+m_2^2\mp\sqrt{(m_2^2-m_1^2)^2+4m_{12}^2}}.
\end{equation}
Here, as in section II, the parameters of the Lagrangian are identified as the physical ones evaluated at the electroweak scale. If we demand $\mu^2>0$, then only SUSY is broken at tree 
level -leaving electroweak symmetry untouched- and from Eq. \refeq{mum} we have $m_1^2m_2^2>m_{12}^4$. 
Also, if we require the potential to be bounded from below, the parameters are constrained to 
satisfy $m_1^2+m_2^2\geq 2m_{12}^2$. 
\bigskip

As usual in this context, we work in the decoupling limit where all SUSY partners of SM particles 
and all physical scalars that emerge from the Higgs sector (except for $\phi$) are heavy, with masses 
of the order of the global SUSY breaking scale $M_S$. From the analysis of the previous section we know 
that the contribution of gauge bosons and fourth generation leptons to the one-loop effective potential 
are negligible; in the first case because of the relative smallness of the gauge couplings compared to 
the quark Yukawa couplings and in the second case because the number of degrees of freedom per lepton is 
$1/3$ that of quarks.  For simplicity, we also discard terms of the form $\phi\varphi^3$, $\phi^2\varphi^2$ 
and $\phi^3\varphi$ because their contribution is also dictated by the gauge couplings.  Under these 
simplifications, the resulting effective potential is given by Eq.\refeq{Vef} with
 $a=t,u_4,d_4,\tilde{t}^{1,2},\tilde{u}_4^{1,2},\tilde{d}_4^{1,2}$ and field-dependent masses
 $m_t^2(\phi)=g_t^2\sin^2\beta\phi^2/2$, $m_{u_4}^2(\phi)=g_{u_4}^2\sin^2\beta\phi^2/2$,
 $m_{d_4}^2(\phi)=g_{d_4}^2\cos^2\beta\phi^2/2$,
\begin{equation}\label{mixmass}
 m_{\tilde{q}^{1,2}}^2(\phi)=\frac{1}{2}\llav{m_{\tilde{q}^L}^2(\phi)+m_{\tilde{q}^R}^2(\phi)\mp
\sqrt{\pacua{m_{\tilde{q}^L}^2(\phi)-m_{\tilde{q}^R}^2(\phi)}^2+4\tilde{A}_q^2m_q^2(\phi)}},
\end{equation}
where $q=t,u_4,d_4$. In the above expression we have
\begin{equation}\label{mixm2}
\begin{array}{ccc}
m^2_{\tilde{t}^L}(\phi)= m^2_{Q_3}+m^2_{t}(\phi) + D^2_{\tilde{t}^L}(\phi), &\qquad& m^2_{\tilde{t}^R}(\phi)= 
m^2_{U_3}+m^2_{t}(\phi) + D^2_{\tilde{t}^R}(\phi), \\ \\
m^2_{\tilde{u}_4^L}(\phi)= m^2_{Q_4}+m^2_{u_4}(\phi) + D^2_{\tilde{u}_4^L}(\phi), &\qquad& m^2_{\tilde{u}_4^R}(\phi)=
 m^2_{U_4}+m^2_{u_4}(\phi) + D^2_{\tilde{u}_4^R}(\phi), \\ \\
m^2_{\tilde{d}_4^L}(\phi)= m^2_{Q_4}+m^2_{d_4}(\phi) + D^2_{\tilde{u}_4^L}(\phi), &\qquad& m^2_{\tilde{d}_4^R}(\phi)=
 m^2_{D_4}+m^2_{d_4}(\phi) + D^2_{\tilde{d}_4^R}(\phi),
\end{array}
\end{equation}
with $m^2_{Q_3},m^2_{U_3},m^2_{D_3}, m^2_{Q_4},m^2_{U_4},m^2_{D_4}$ as soft supersymmetry-breaking mass 
parameters for the left- and right-handed squarks and
\begin{equation}
\begin{array}{ccc}
D^2_{\tilde{q}^L}(\phi)=m_Z^2(\phi)\cos2\beta\pacua{T_{3L}(\tilde{q})-Q(\tilde{q})\sin^2\theta_W}, &
\qquad& D^2_{\tilde{q}^R}(\phi)=m_Z^2(\phi)\cos2\beta Q(\tilde{q})\sin^2\theta_W.
\end{array}
\end{equation}
Note that the discussion about Yukawa couplings given in the previous section applies in this case to 
the quantities $g_t^{*}=g_t\sin\beta$, $g_{u_4}^{*}=g_{u_4}\sin\beta$ and $g_{d_4}^{*}=g_{d_4}\cos\beta$ for fixed $\beta$. 
In Eq.\refeq{mixmass}, the parameters $\tilde{A}_q$ control the mixing between squarks in each generation. 
We assume that there is no mixing, taking $\tilde{A}_t=\tilde{A}_{u_4}=\tilde{A}_{d_4}=0$. The degrees of freedom
 per particle are $n_{\tilde{q}^{1}}=n_{\tilde{q}^{2}}=6$, $n_q=-12$. Notice also that in this case the tree-level scalar self interactions cannot be taken as zero.
\bigskip

The parameter $\mu$ can be expressed now in terms of the physical masses after the minimization of 
the effective potential. At $\phi=v$ one obtains
\begin{equation}
 \mu^2=-\frac{1}{2}m_Z^2\cos^2 2\beta+\sum_{q}\frac{3 m_q^2}{8\pi^2 v^2}\pacua{m_{\tilde{q}^1}^2
\ln\paren{1+\frac{\Lambda^2}{m_{\tilde{q}^1}^2}}+
m_{\tilde{q}^2}^2\ln\paren{1+\frac{\Lambda^2}{m_{\tilde{q}^2}^2}}-2 m_q^2\ln\paren{1+\frac{\Lambda^2}{m_q^2}}},
\end{equation}
with $\mu^2>0$ again for the present set up. For the Higgs mass and the effective Higgs self-coupling 
at electroweak scale one has
\begin{equation}\label{HMs}
\left.\frac{\partial^2 V^{(1)}}{\partial\phi^2}\right|_{\phi=v}=m_H^2=m_Z^2\cos^2 2\beta 
+\sum_{q}\frac{3 m_q^4}{4\pi^2v^2}\ln\paren{\frac{m_{\tilde{q}^1}^2 m_{\tilde{q}^2}^2}{m_{q}^4}}
\end{equation}
and 
\begin{equation}\label{cuartas}
\begin{split}
\left.\frac{\partial^4 V^{(1)}}{\partial\phi^4}\right|_{\phi=v}=&\frac{3m_Z^2\cos^2 2\beta}{v^2}
\\&+\sum_{q}\frac{3  m_q^4}{4\pi^2v^4}\llav{3\ln\paren{\frac{m_{\tilde{q}^1}^2 m_{\tilde{q}^2}^2}{m_{q}^4}}
-4\pacua{m_q^4\paren{\frac{1}{m_{\tilde{q}^1}^4}+\frac{1}{m_{\tilde{q}^2}^4}}-3m_q^2\paren{\frac{1}{m_{\tilde{q}^1}^2}
+\frac{1}{m_{\tilde{q}^2}^2}}+4}}                               
\end{split}
\end{equation}
neglecting terms that vanish as $\Lambda\to\infty$ because the soft breaking terms for the squarks play the 
role of natural regulators in this case.
\bigskip

\begin{center}
\begin{figure}[ptb]
\begin{center}
\includegraphics[width=0.8 \textwidth]{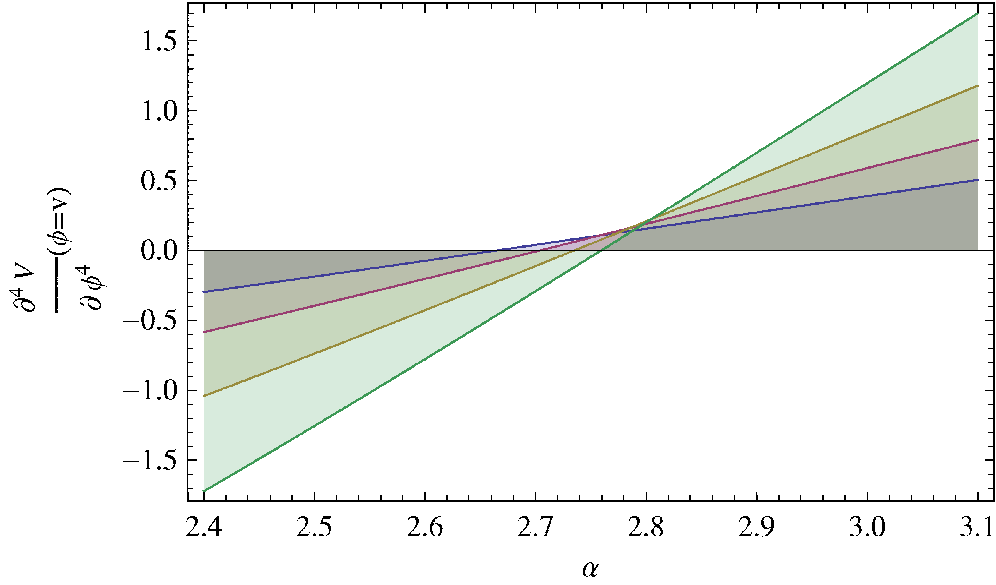}
\end{center}
\caption{Effective Higgs self-coupling at the electroweak scale $v$ as a function of $\alpha$ for fixed 
values of the heavy fermion masses and $\beta=\pi/4$. The curves correspond to $m_{u_4}=$350, 400, 450 and 500 GeV  
from left to right at zero.}
\label{lambdams}
\end{figure}
\end{center}

\begin{center}
\begin{figure}[ptb]
\begin{center}
\includegraphics[width=0.8 \textwidth]{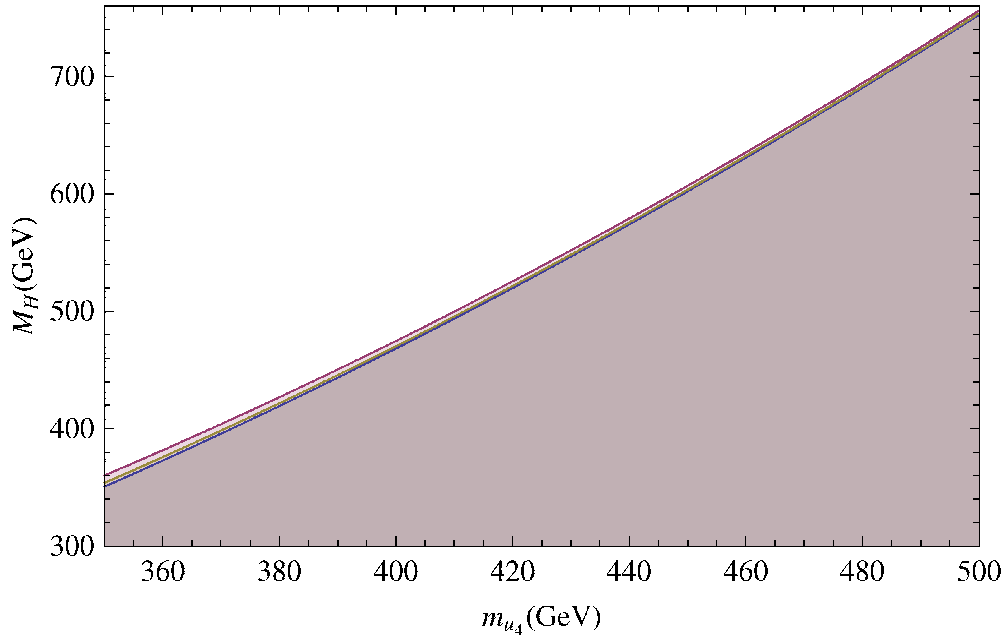}
\end{center}
\caption{Higgs mass as a function of $m_{u_4}$. The external lines correspond to the extreme cases 
$\beta=\pi/2$ (bottom) and $\beta=\pi/4$ (top). The middle line contains only the contribution of 
$u_4$, $d_4$ and their super-partners with $\beta=\pi/4$.}
\label{mums}
\end{figure}
\end{center}

Again, if we insist that electroweak SB is completely produced by quark and squark loops, consistency 
requires that Higgs self interactions must remain small at least at the scale of SB as in Eq.\refeq{cuarta0}. 
Taking for all squarks the same soft mass $m_s=m_{Q_3}=m_{U_3}= m_{Q_4}=m_{U_4}=m_{D_4}=\alpha m_{u_4}\sim M_{S}$ 
in Eq.\refeq{mixm2}, one can extract the maximum value of $\alpha$ allowed by 
$\left.{\partial^4 V^{(1)}}/{\partial\phi^4}\right|_{\phi=v}=0$. This is shown in Fig. (\ref{lambdams}). 
The solution turns to be very stable and lies in the range 
\begin{equation}\label{rangea}
\begin{array}{ccc}
2.3<\alpha<2.8 &\Rightarrow \qquad& 800\text{ GeV}<m_s<1400\text{ GeV},
\end{array}
\end{equation}
for $0\le\beta\le\pi/2$ with fixed values of $g_t^*$, $g_{u_4}^*$ and $g_{d_4}^*$ and the relation 
Eq.\refeq{dif} for the masses of the fourth generation quarks. The results turn to be weakly 
$\beta$ dependent as we will see in Fig. (\ref{mums}).	

Finally, once determined the parameter $\alpha$, Eq.\refeq{HMs} leads to the corresponding Higgs mass 
upper bound in the limit $\Lambda\to\infty$ as a function of $\beta$ and $m_{u_4}$ as shown in 
Fig. (\ref{mums}). The prediction for the Higgs mass is very similar to that of the previous section, 
from $350 $ GeV to about $750$ GeV up to small corrections that would come from gauge bosons, leptons 
and sleptons in the loop, which are expected to modify our results only a few percent. In fact, even 
the contribution of top quarks is negligible (see Fig. (\ref{mums})). From Eq.\refeq{HMs}, Eq.\refeq{cuartas} 
and Eq.\refeq{cuarta0}, the dominant contribution to Higgs mass (taking $\beta=\pi/4$ for simplicity, 
which implies $m_{\tilde{q}}^2=m_{\tilde{q}^1}^2=m_{\tilde{q}^2}^2=m_s^2+m_q^2$) is:
\begin{equation}\label{HM2}
m_H^2\approx\sum_{q=u_4,d_4}\frac{4 m_q^4}{\pi^2v^2}\pacua{1-\frac{3}{2}\frac{m_q^2}{m_{\tilde{q}}^2}
+\frac{1}{2}\frac{m_q^4}{m_{\tilde{q}}^4}},
\end{equation}
with $\alpha$ and $m_s$ given by Eq.\refeq{rangea}. Given the small difference between $m_{u_4}$ and 
$m_{d_4}$, a good approximation to Eq.\refeq{HM2} is simply 
\begin{equation}\label{HM3}
m_H\approx\frac{1.83}{\pi v}\sqrt{m_{u_4}^4+m_{d_4}^4},
\end{equation}
which correspond to $\alpha\approx 2.8$ and $980 \text{ GeV}< m_s < 1400 \text{ GeV }$.

\section{Conclusions}
In this paper, we study the possibility of electroweak symmetry breaking by radiative corrections 
\cite{Coleman:1973jx} due to a fourth generation in the Standard Model. We isolate the effects of the 
fourth generation  by taking a vanishing scalar self-coupling at the classical level and maintaining 
this condition valid at one loop level at the electroweak symmetry breaking scale. 
In such a scenario, electroweak symmetry is broken by radiative corrections due mainly to the fourth 
generation and Higgs masses of the order of a few hundreds of GeV are consistent with electroweak precision 
data. Furthermore, the theory is valid only up to a scale $\Lambda \sim 1-2$ TeV. Such low cut-off means that 
the effects of new physics needed to describe electroweak interactions at energy above $\Lambda$ should be 
measurable at the LHC.  We use the renormalization group equation to study the impact of the running of the 
Yukawa couplings in our results. We show that the predictions of the model are not strongly modified 
by the running of the Yukawa couplings, but a slightly lower cut-off related to the breaking of the perturbative 
regime is expected in this case.  

As an example of models with new physics and therefore containing a natural scale for the cut-off of 
the electroweak interactions regime, we study a simplified Minimal Supersymmetric Standard Model with four 
generations. We obtain similar values for the Higgs mass with weak $\beta$ dependence. The natural scale for 
the cut-off of the electroweak regime is given by the mass of the fourth generation squarks and 
EWSB by radiative corrections due predominantly to the fourth generation  
requires masses for the squarks of the order $m_s\sim 1 \; \mathrm{TeV}$.
 
\section{acknowledgements}
This work was supported by CONACyT (Mexico) under grant 50471-F, DINPO-UG and PROMEP-SEP.

\end{document}